\documentclass[12pt]{article}
\usepackage[english]{babel}
\usepackage{authblk}
\usepackage{graphicx}
\usepackage{subcaption}
\usepackage{amssymb}
\usepackage{float}
\usepackage{amssymb}
\usepackage{amsmath}
\usepackage{wrapfig}
\usepackage[left=20mm, top=15mm, right=15mm, bottom=15mm, nohead, footskip=10mm]{geometry}

\title{Quantum "Twin Peaks" or Path Integrals in the Future Light Cone.}

\author[a]{Vladimir V. Belokurov\thanks{vvbelokurov@yandex.ru}} 
\author[a]{Vsevolod V. Chistiakov\thanks{vsevolod.chistyakov@gmail.com}}
\author[a]{Klavdiia A. Lursmanashvili\thanks{klavo4ka2001@gmail.com}}
\author[a]{Evgeniy T. Shavgulidze\thanks{shavgulidze@bk.ru}}

\affil[a]{Lomonosov Moscow State University.}

\date{}

\begin{document}
\maketitle
\abstract
{By analogy with
the Wiener measure on the Euclidean plane that is invariant under the group of rotations and quasi-invariant under the group of diffeomorphisms,
we construct the path integrals measure that is invariant under the Lorentz group and quasi-invariant under the group of diffeomorphisms.

The correspondence between the paths in the future cone of the Minkowskian plane and the paths in the coverings of the Euclidean plane is established.}

\section{Introduction}

In quantum field theory and in the theory of gravity, there exists a problem of analytically continuing the spacetime metric in order to properly define the corresponding path integrals \cite{Konts}, \cite{Witten}. Consider a path integral of the form
\begin{equation}\label{rel PI}
    K(x,y,T)=\int \delta(\xi(0)-x)\,\delta(\xi(T)-y)\,\exp\left(-\frac{1}{2}\int_0^T d\tau \,\,\left[(\dot \xi^0)^2-(\dot \xi^1)^2\right]\right)\,d\xi
\end{equation}
arising \cite{Kaku} in the description of the propagation of a relativistic particle from a point $x$ to a point $y$ in spacetime. Formally, the integral of the form (\ref{rel PI}) does not exist due to the presence of a factor $\sim\exp\left(+\frac{1}{2}\int d\tau\, (\dot \xi^1)^2\right)$ that grows at infinity. The standard technique consists in considering the integrals of the form
\begin{equation}\label{alphaint}
      K_{\alpha}(x,y,T)=\int \delta(\xi(0)-x)\,\delta(\xi(T)-y)\,\exp\left(-\frac{1}{2}\int_0^T d\tau \,\,\left[(\dot \xi^0)^2+\alpha(\dot \xi^1)^2\right]\right)\,d\xi\, ,
\end{equation}
depending on a complex parameter $\alpha$. For positive real $\alpha$, the integrals (\ref{alphaint}) represent the integrals with over the Wiener measure \cite{Shyr}. The integral of the form (\ref{rel PI}) is postulated to be the integral (\ref{alphaint}), evaluated for $\alpha>0$, and then analytically continued to $\alpha= -1$.

In this paper using the decomposition of the Wiener measure over the orbits of the group of diffeomorphisms \cite{unusual view}-\cite{polar2d} we construct the measure on the space of trajectories lying in the future cone of the Minkowskian plane, that is invariant under the Lorentz transformations and quasi-invariant under the group of diffeomorphisms.  

In Sections 2-3, we recall some known properties of the Wiener measure on trajectories in one-dimensional and two-dimensional Euclidean space. In Section 4, we construct a measure on trajectories lying in the future cone. In Section 5, we discuss the isometric mapping between the cone and the infinite-sheeted covering of the Euclidean plane. In Section 6, we discuss the obtained results and some of their possible applications. Appendix A provides the explicit expressions for the mappings used to formulate the causality properties. Appendix B demonstrates the equivalence of the forms of the obtained measure in Cartesian and decomposition coordinates.

\section{The Wiener measure on trajectories lying on $\mathbb{R}$}
Consider the Wiener measure on the space of continuous functions $C([0,T])$
\begin{equation}\label{win1d}
    w_{\sigma}(d\xi)=\exp\left(-\frac{1}{2\sigma^2}\int_0^T \dot \xi^2(\tau) \,d\tau\right)d\xi\, .
\end{equation}
We suppose the left endpoint of the random process $\xi(\tau)$ to be not fixed.

The Wiener measure on the space $C([0,T])$ is trivially transferred to the measure on the space $C([t_0,t_0+T])$. Namely, let $X'$ be a Borel subset of $C([t_0,t_0+T])$, and $w'_{\sigma}$ be the Wiener measure on $C([t_0,t_0+T])$. We define the argument shift transformation $\mathcal{T}_{t_0}:C([t_0,t_0+T])\to C([0,T])$ as
\begin{equation}
    (\mathcal{T}_{t_0}\xi)(\tau)\equiv\xi(\tau+t_0).
\end{equation}
Applying this transformation to each element $\xi\in X'$, we obtain the set
\begin{equation}
    \mathcal{T}_{t_0}X'\equiv\{\mathcal{T}_{t_0}\xi\,;\,\xi\in X'\}\,, \qquad \mathcal{T}_{t_0}X'\subset C([0,T]) .
\end{equation}
The Wiener measure of any Borel subset does not change under an argument shift
\begin{equation}
    w'_{\sigma}(X')=w_{\sigma}(\mathcal{T}_{t_0}X')\, .
\end{equation}

In the theory of stochastic processes \cite{Shyr}, it is proved that for any $t^{*}\in[0,T]$, the values of the Wiener process $\xi(\tau)$ on the interval $[t^{*},T]$ are independent of the values of the same process on the interval $[0,t^{*})$, and depend only on $\xi^{*}=\xi(t^{*})$.

Consider the set $\mathcal{X}\subset C([0,t^{*}])\times C([t^{*},T])$ of the form
\begin{equation}
    \mathcal{X}=\{(\xi_1,\xi_2)\in C([0,t^{*}])\times C([t^{*},T])\,;\,\xi_1(t^{*})=\xi_2(t^{*})\}\, .
\end{equation}
We construct a one-to-one correspondence $\Xi:\mathcal{X}\to C([0,T])$ as follows
\begin{equation}\label{omega}
    \xi=\Xi(\xi_1,\xi_2)\equiv \left\{\begin{array}{cc}
        \xi_1(\tau) &  \tau\in[0,t^{*}]\\
        \xi_2(\tau) &  \tau\in [t^{*},T]\, .
    \end{array}\right.\
\end{equation}
In other words, we split the process $\xi(\tau)$ on the interval into the two subprocesses on the intervals $[0,t^{*}]$ and $[t^{*},T]$, respectively.

Let an arbitrary functional $F(\xi)$ on the space $C([0,T])$ be given. We construct a functional $\tilde F(\xi_1,\xi_2)$ on the space $C([0,t^{*}])\times C([t^{*},T])$ as follows. If $(\xi_1,\xi_2)\in \mathcal{X}$, we set by definition
\begin{equation}
    \tilde F(\xi_1,\xi_2)=F(\Xi(\xi_1,\xi_2))\, .
\end{equation}
We extend $\tilde F$ to the space $C([0,t^{*}])\times C([t^{*},T])$ by requiring the functional to be invariant under a shift of the second argument by a constant
\begin{equation}
    \tilde F(\xi_1,\xi_2+c)=\tilde F(\xi_1,\xi_2)\, .
\end{equation}
Let $w^{1}_{\sigma}$ be the Wiener measure on $C([0,t^{*}])$, and $w^{2}_{\sigma}$ be the Wiener measure on $C([t^{*},T])$. Then
\begin{equation}\label{mark}
    \int F(\xi)\,w_{\sigma}(d\xi)=\int \tilde  F(\xi_1,\xi_2)\,\delta(\xi_1(t^{*})-\xi_2(t^{*}))\,w^1_{\sigma}(d\xi_1)\,w^2_{\sigma}(d\xi_2)\, .
\end{equation}

{\bf Quasi-invariance with respect to the groups of diffeomorphisms.}\\
Let the action of the group of the diffeomorphisms $Diff^1_{+}[0,T]$  on the space $C([0,T])$ is given by\footnote{Here $\dot \varphi^{-1}(\tau)\equiv\frac{d}{d\tau}\varphi^{-1}(\tau)$}
\begin{equation}\label{diffact}
    (\varphi \xi)(\tau)=\frac{\xi(\varphi^{-1}(\tau))}{\sqrt{\dot\varphi^{-1}(\tau)}}\, , \qquad \varphi\in Diff^1_{+}[0,T]\, .
\end{equation}
The Wiener measure is quasi-invariant under the action of the group $Diff^1_{+}([0,T])$. That is \cite{Shepp}
\begin{equation}
    w_{\sigma}(d(\varphi\xi))=\mathcal{P}_{\varphi}(\xi)w_{\sigma}(d\xi)\, .
\end{equation}
Here,
\begin{equation}
    \mathcal{P}_{\varphi}(\xi)=\sqrt[4]{\dot\varphi(0)\dot \varphi(T)}\,\exp\left(\frac14\left[(\xi(T))^2\frac{\ddot \varphi(T)}{\dot \varphi(T)}-(\xi(0))^2\frac{\ddot \varphi(0)}{\dot \varphi(0)}\right]\right)\,\exp\left(\frac14\int_0^Td\tau \,\,(\xi(\tau))^2\mathcal{S}ch\{\varphi,\tau\}\right)\, 
\end{equation}
is the Radon-Nikodym derivative, and
\begin{equation}
    \mathcal{S}ch\{\varphi,\tau\}\equiv\frac{d}{d\tau}\left(\frac{\ddot \varphi}{\dot \varphi}\right)-\frac12\left(\frac{\ddot \varphi}{\dot \varphi}\right)^2\, 
\end{equation}
is the Schwarzian derivative.

{\bf Decomposition of the Wiener measure over the orbits of the $Diff^1_{+}([0,T])$}\\
Let $C_{+}([0,T])\subset C([0,T])$ be the space of continuous positive functions on the interval $[0,T]$. $C_{+}([0,T])$ is decomposed into the orbits of the action (\ref{diffact}) of the group $Diff^1_{+}([0,T])$ \cite{unusual view}, \cite{polar}, \cite{polar2d} with definite value of the group invariant
\begin{equation}
    \frac{1}{\rho^{2}}\equiv\frac{1}{T}\int_0^T \frac{d\tau}{\xi^2(\tau)}\, .
\end{equation}

Using the decomposition of the space $C_{+}[0,T]$ into the orbits of the action of the group $Diff_{+}^1([0,T])$, we represent the function $\xi(\tau)\in C_{+}([0,T])$ as follows. On every orbit with the definite $\rho$, we consider the constant function $\tilde \xi_{\rho}(\tau)=\rho$. All other elements of the orbit are obtained by applying the action (\ref{diffact}) of the group of diffeomorphisms on $\tilde \xi_{\rho}$. Thus, any element $\xi(\tau)\in C_{+}[0,T]$ can be represented in the form \cite{polar}
\begin{equation}
    \xi(\tau)=(\varphi\tilde\xi_{\rho})(\tau)=\rho\frac{1}{\sqrt{\dot\varphi^{-1}(\tau)}}\, .
\end{equation}
To every element $\xi\in C_{+}([0,T])$, we uniquely associate a pair $(\rho,\varphi)$ of decomposition coordinates, where $\rho\in \mathbb{R}^{+}$, $\varphi\in Diff^1_{+}([0,T])$ according to the rule
\begin{equation}
    \frac{1}{\rho^2}=\frac{1}{T}\int_0^T\frac{d\tau_1}{\xi^2(\tau_1)}\, ,\qquad \varphi^{-1}(\tau)=\frac{\int_0^{\tau}\frac{d\tau_1}{\xi^2(\tau_1)}}{\frac{1}{T}\int_0^{T}\frac{d\tau_1}{\xi^2(\tau_1)}}\, .
\end{equation}

The Wiener measure on the space $C_{+}[0,T]$ can be represented as a decomposition \cite{polar} over the orbits of the action of the group $Diff^1_{+}([0,T])$
\begin{equation}
    w_{\sigma}(d\xi)=\exp\left(-\frac{\sigma^2}{4\rho^2}\right)\,(\dot \varphi(0)\dot \varphi(T))^{3/4}\,\mu_{\frac{2\sigma}{\rho}}(d\varphi)\,d\rho\, ,
\end{equation}
where $\mu$ is the quasi-invariant measure on the group of diffeomorphisms \cite{Shavg}, \cite{calc}
\begin{equation}\label{mu}
    \mu_{\sigma}(d\varphi)=\frac{1}{\sqrt{\dot \varphi(0)\dot \varphi(T)}}\exp\left\{\frac{1}{\sigma^2}\left[\frac{\ddot \varphi(0)}{\dot \varphi(0)}-\frac{\ddot \varphi(T)}{\dot \varphi(T)}\right]\right\}\exp\left\{\frac{1}{\sigma^2}\int_0^T\mathcal{S}ch\{\varphi,\tau\}\, \, d\tau\right\}d\varphi\, .
\end{equation}
The symbol $d\varphi$ here denotes the generalized Haar measure on the group $Diff_{+}^1([0,T])$
\begin{equation}
    d\varphi\equiv\prod_{\tau_i\in[0,T]}\frac{d\varphi(\tau_i)}{\dot \varphi(\tau_i)}\, .
\end{equation}

Let us rewrite (\ref{mark}) in the decomposition coordinates. As in equ. (\ref{mark}), we split the random process on the interval $[0,T]$ into the two subprocesses on the intervals $[0,t^{*}]$ and $[t^{*},T]$. We write each subprocess in decomposition coordinates
\begin{equation}
    \xi_1(\tau)=\rho_1\frac{1}{\sqrt{\dot \varphi^{-1}_{1}(\tau)}}\, ,\qquad \xi_2(\tau)=\rho_2\frac{1}{\sqrt{\dot\varphi_2^{-1}(\tau)}}\, ,
\end{equation}
where $\varphi_1\in Diff_{+}^1([0,t^{*}])$, $\varphi_2\in Diff^1_{+}([t^{*},T])$.

The decomposition coordinates $(\rho_1,\varphi_1)$ and $(\rho_2,\varphi_2)$ are expressed in terms of the decomposition coordinates $(\rho,\varphi)$ of the original process $\xi(\tau)$ using a one-to-one mapping
\begin{equation}\label{OmegaR}
    (\rho,\varphi)=\Omega(\rho_1,\varphi_1,\rho_2,\varphi_2)\, .
\end{equation}
The explicit forms of the mapping $\Omega$ and the inverse mapping $\Omega^{-1}$ are given in Appendix A.

Let $\Phi(\rho,\varphi)$ be a functional  on $\mathbb{R}^{+}\times Diff_{+}^1$. We construct a functional $\tilde \Phi(\rho_1,\varphi_1,\rho_2,\varphi_2)$ on $\mathbb{R}^{+}\times Diff_{+}^1([0,t^{*}])\times\mathbb{R}^{+}\times Diff_{+}^1([t^{*},T])$ according to the rule
\begin{equation}
    \tilde \Phi(\rho_1,\varphi_1,\rho_2,\varphi_2)=\Phi(\Omega(\rho_1,\varphi_2,\rho_2,\varphi_2)).
\end{equation}
Let $\mu^1_{\sigma}$ be the quasi-invariant measure (\ref{mu}) on the group $Diff([0,t^{*}])$, and $\mu^2_{\sigma}$ be the quasi-invariant measure on the group $Diff_{+}^1([t^{*},T])$. The following equality is valid
\begin{equation}\label{1ddecomp}
    \int \Phi(\rho,\varphi)\,\exp\left(-\frac{\sigma^2}{4\rho^2}\right)\,(\dot \varphi(0)\dot \varphi(T))^{3/4}\,\mu_{\frac{2\sigma}{\rho}}(d\varphi)\,d\rho=
\end{equation}
\begin{equation*}
    =\int \tilde \Phi(\rho_1,\varphi_1,\rho_2,\varphi_2)\,\delta\left(\rho_1\sqrt{\dot \varphi_1(t^{*})}-\rho_2\sqrt{\dot \varphi_2(t^{*})}\right)\,\exp\left(-\frac{\sigma^2}{4\rho_1^2}-\frac{\sigma^2}{4\rho_2^2}\right)
\end{equation*}
\begin{equation*}
    (\dot \varphi_1(0)\dot \varphi_1(t^{*})\dot \varphi_2(t^{*})\dot \varphi_2(T))^{3/4}\, \mu^1_{\frac{2\sigma}{\rho_1}}(d\varphi_1)\,\mu^2_{\frac{2\sigma}{\rho_2}}(d\varphi_2)\,d\rho_1\,d\rho_2\, .
\end{equation*}

\section{The Wiener measure on trajectories in the Euclidean plane}
Now consider the space $C([0,T],\mathbb{R}^2)$ of the continuous functions $\xi(\tau)=(\xi^0(\tau),\xi^1(\tau))$ defined on the $[0,T]$, with the values in $\mathbb{R}^2$. On this space, the Wiener measure is defined as follows
\begin{equation}\label{win2d}
    w_{\sigma}(d\xi)=w_{\sigma}(d\xi^0)\,w_{\sigma}(d\xi^1)=\exp\left(-\frac{1}{2\sigma^2}\int_0^Td\tau\left[({\dot \xi^0}(\tau))^2+({\dot \xi^1}(\tau))^2\right]\right)d\xi^0\,d\xi^1\, .
\end{equation}

The measure (\ref{win2d}) is invariant under the group $SO(2)$
\begin{equation}\label{Rinv}
    w_{\sigma}(d(R_{\gamma}\xi))=w_{\sigma}(d\xi)\, ,
\end{equation}
where
\begin{equation}\label{rot}
    R_{\gamma}\in SO(2)\, ,\qquad R_{\gamma}\xi(\tau)\equiv\left(\xi^0(\tau)\cos\gamma-\xi^1(\tau) \sin\gamma,\,\xi^0(\tau)\sin\gamma+\xi^1(\tau)\cos\gamma \right)\, .
\end{equation}

Similarly to the Wiener measure on $C([0,T])$, the measure (\ref{win2d}) possesses the properties of causality and quasi-invariance with respect to the action of the group $Diff^1_{+}([0,T])$. To define the action of the group of the diffeomorphisms $Diff_{+}^1([0,T])$ on $C([0,T],\mathbb{R}^2)$ it is convenient to use the polar coordinates defined by
\begin{equation}
    \xi^0(\tau)=r(\tau)\cos\theta(\tau)\,, \qquad \xi^1(\tau)=r(\tau)\sin\theta(\tau)\, .
\end{equation}
The group $Diff_{+}^1([0,T])$ acts as
\begin{equation}\label{act2diff}
    (\varphi \xi)(\tau)=\left(\frac{r(\varphi^{-1}(\tau))}{\sqrt{\dot \varphi^{-1}(\tau)}}\cos\theta(\varphi^{-1}(\tau)),\frac{r(\varphi^{-1}(\tau))}{\sqrt{\dot \varphi^{-1}(\tau)}}\sin\theta(\varphi^{-1}(\tau))\right)\, .
\end{equation}
Note that the action of the group $Diff_{+}^1([0,T])$ defined by (\ref{act2diff}) commutes with the action of the rotation group
\begin{equation}
 R_{\gamma}\varphi \xi=\varphi R_{\gamma}\xi   
\end{equation}

The invariant of the action of the group of diffeomorphisms (\ref{act2diff}) has the form
\begin{equation}
\frac{1}{\rho^{2}}=\frac{1}{T}\int_0^T  \frac{d\tau}{(\xi^0(\tau))^2+(\xi^1(\tau))^2}\, .
\end{equation}

For each orbit of the action of the group $Diff^1_{+}([0,T])$ consider one element of the form \cite{polar2d}
\begin{equation}\label{one element}
    \tilde \xi_{\rho,\psi,\alpha}(\tau)=(\rho\cos\left[\psi(\tau)+\alpha\right],\rho\sin\left[\psi(\tau)+\alpha\right])\, ,
\end{equation}
where $\psi(\tau)\in C_0([0,T])$, $\alpha\in S^1$. Here, $C_{0}([t_0,t_0+T])$ is the space of continuous functions with a fixed left endpoint $\xi(t_0)=0$ and $S^1=\mathbb{R}/2\pi\mathbb{Z}$ is the circle.

An arbitrary function $\xi\in C([0,T],\mathbb{R}^2)$ can be obtained from a function of the form (\ref{one element}) by equ. (\ref{act2diff})
\begin{equation}
    \xi=(\varphi\tilde \xi_{\rho,\psi,\alpha})(\tau)=\left(\frac{\rho}{\sqrt{\dot \varphi^{-1}(\tau)}}\cos\left[\psi(\varphi^{-1}(\tau))+\alpha\right]\, ,\frac{\rho}{\sqrt{\dot\varphi^{-1}(\tau)}}\sin\left[\psi(\varphi^{-1}(\tau))+\alpha\right]\,\right) .
\end{equation}
Thus, there is one to one correspondence
\begin{equation}
    \xi\longleftrightarrow(\rho,\psi,\varphi,\alpha)\, ,
\end{equation}
where $\xi(\tau)\in C([0,T],\mathbb{R}^2)$,\, $\rho\in \mathbb{R}^{+}$,\, $\psi\in C_0([0,T])$,\, $\varphi\in Diff^1_{+}([0,T])$,\, $\alpha\in S^1$
\begin{equation}
    \frac{1}{\rho^2}=\frac{1}{T}\int_0^T\frac{d\tau_1}{(\xi^0(\tau_1))^2+(\xi^1(\tau_1))^2}\, ,\qquad \varphi^{-1}(\tau)=\frac{\int_0^{\tau}\frac{d\tau_1}{(\xi^0(\tau_1))^2+(\xi^1(\tau_1))^2}}{\frac{1}{T}\int_0^T\frac{d\tau_1}{(\xi^0(\tau_1))^2+(\xi^1(\tau_1))^2}}\, ,
\end{equation}
\begin{equation}
     \psi(s)=\mathrm{arg} \,z(\varphi(s))-\mathrm{arg}\,z(0)\, , \qquad \alpha=\mathrm{Arg}\, z(0)\, ,
\end{equation}
where
\begin{equation}
    z(\tau)=\xi^0(\tau)+i\xi^1(\tau)\,.
\end{equation}

The decomposition of the Wiener measure over the orbits of the action of the group $Diff_{+}^1([0,T])$ (\ref{act2diff}) has the form \cite{polar2d}
\begin{equation}\label{e2dpolar}
    w_{\sigma}(d\xi)=\rho\exp\left(-\frac{\sigma^2}{4\rho^2}\right)\,\dot \varphi(0)\dot \varphi(T)\, \mu_{\frac{2\sigma}{\rho}}(d\varphi)\,w^{0}_{\frac{\sigma}{\rho}}(d\psi)\,d\rho\, d\alpha\, .
\end{equation}
Here, $w^0_{\sigma}$ is the Wiener measure on the space $C_0([0,T])$.

The group $SO(2)$ acts on decomposition coordinates as follows
\begin{equation}\label{gamma decomp}
    R_{\gamma}(\rho,\psi,\varphi,\alpha)\longleftrightarrow(\rho,\psi,\varphi,\alpha+\gamma)\, .
\end{equation}
The invariance of the measure (\ref{e2dpolar}) is obvious.

In the two-dimensional case, there is an identity analogous to (\ref{1ddecomp}). We split the random process $\xi\in C([0,T],\mathbb{R}^2)$ into two subprocesses: let $\xi_1(\tau)$ coincide with $\xi(\tau)$ on the interval $[0,t^{*}]$, and $\xi_2(\tau)$ coincide with $\xi(\tau)$ on the interval $[t^{*},T]$. To each of the subprocesses we associate a set of decomposition coordinates
\begin{equation}
    \xi_1\longleftrightarrow(\rho_1,\psi_1,\varphi_1,\alpha_1)\, ,\qquad \xi_2\longleftrightarrow(\rho_2,\psi_2,\varphi_2,\alpha_2)
\end{equation}
The coordinates $(\rho_1,\psi_1,\varphi_1,\alpha_1)$ and $(\rho_2,\psi_2,\varphi_2,\alpha_2)$ are uniquely expressed in terms of the coordinates $(\rho,\psi,\varphi,\alpha)$ of the original process
\begin{equation}\label{OmegaR2}
    (\rho,\psi,\varphi,\alpha)=\Omega^{\mathbb{R}^2}(\rho_1,\psi_1,\varphi_1,\alpha_1,\rho_2,\psi_2,\varphi_2,\alpha_2)
\end{equation}
The explicit form of $\Omega^{\mathbb{R}^2}$ is given in Appendix A. Let $\Phi(\rho,\psi,\varphi,\alpha)$ be an arbitrary functional on $\mathbb{R}^{+}\times C_0([0,T])\times Diff_{+}^{1}([0,T])\times S^1$. We construct a functional $\tilde \Phi(\rho_1,\psi_1,\varphi_1,\alpha_1,\rho_2,\psi_2,\varphi_2,\alpha_2)$ as follows
\begin{equation}
    \tilde \Phi(\rho_1,\psi_1,\varphi_1,\alpha_1,\rho_2,\psi_2,\varphi_2,\alpha_2)=\Phi(\Omega^{\mathbb{R}^2}(\rho_1,\psi_1,\varphi_1,\alpha_1,\rho_2,\psi_2,\varphi_2,\alpha_2))\,.
\end{equation}
Let $\mu^1_{\sigma}$ be the quasi-invariant measure on $Diff^1_{+}([0,t^{*}])$, $\mu^2_{\sigma}$ be the quasi-invariant measure on $Diff^1_{+}([t^{*},T])$, $w^{01}_{\sigma}$ be the Wiener measure on $C_0([0,t^{*}])$, $w^{02}_{\sigma}$ be the Wiener measure on $C_0([t^{*},T])$. The following equality is valid
\begin{equation}\label{2ddecomp}
    \int \Phi(\rho,\varphi,\psi,\alpha)\,\,\rho \,\exp\left(-\frac{\sigma^2}{4\rho^2}\right)\,\dot \varphi(0)\dot \varphi(T)\, \mu_{\frac{2\sigma}{\rho}}(d\varphi)\,w^{0}_{\frac{\sigma}{\rho}}(d\psi)\,d\rho\, d\alpha=
\end{equation}
\begin{equation*}
    =\int \tilde \Phi(\rho_1,\varphi_1,\psi_1,\alpha_1,\rho_2,\varphi_2,\psi_2,\alpha_2)\frac{\delta\left(\rho_1\sqrt{\dot \varphi_1(t^{*})}-\rho_2\sqrt{\dot \varphi_2(t^{*})}\right)}{\rho_1\sqrt{\dot \varphi_1(t^{*})}}\sum_{n}\delta\left(\alpha_2-\psi_1(t^*)-\alpha_1-2\pi n\right)
\end{equation*}
\begin{equation*}
    \rho_1\rho_2\,\exp\left(-\frac{\sigma^2}{4\rho_1^2}-\frac{\sigma^2}{4\rho_2^2}\right)\,\dot \varphi_1(0)\dot \varphi_1(t^{*})\dot \varphi_2(t^{*})\dot \varphi_2(T)
\end{equation*}
\begin{equation*}
    \mu^1_{\frac{2\sigma}{\rho_1}}\,(d\varphi_1)\mu^{2}_{\frac{2\sigma}{\rho_2}}(d\varphi_2)\,w^{01}_{\frac{\sigma}{\rho_1}}(d\psi_1)\,w^{02}_{\frac{\sigma}{\rho_2}}(d\psi_2)\,d\rho_1\,d\rho_2\,d\alpha_1\,d\alpha_2\, .
\end{equation*}

\section{The measure on trajectories in the future cone}
We define the future cone in the plane as follows
\begin{equation}
    \mathrm{Cone}=\{(x^0,x^1)\in \mathbb{R}^2\, ,\, x^0>0,|x^1|< x^0\}\, , \qquad \mathrm{Cone} \subset \mathbb{R}^2\, .
\end{equation}
Let $C([0,T],\mathrm{Cone})$ be the space of continuous functions $\xi(\tau)$ with values in $\mathrm{Cone}$.
Analogously to how we considered the action of $SO(2)$ on the two-dimensional Euclidean plane, we consider the action of the group $SO(1,1)$ (the Lorentz group) on $\mathrm{Cone}$
\begin{equation}\label{Lorentz}
    L_{\gamma}\xi(\tau)=(\,\xi^0(\tau)\,\cosh \gamma + \xi^1(\tau)\,\sinh\gamma ,\,\xi^0(\tau)\,\sinh\gamma + \xi^1(\tau)\,\cosh\gamma)
\end{equation}

We define Minkowskian coordinates $r,\theta$ by the relations
\begin{equation}
    \xi^0=r(\tau)\cosh(\theta(\tau))\, ,\qquad \xi^1=r(\tau)\sinh(\theta(\tau))\, .
\end{equation}
On the space $C([0,T],\mathrm{Cone})$, we define the action of the diffeomorphism group as follows
\begin{equation}\label{pseudo act}
    (\varphi \xi)(\tau)=\left(\frac{r(\varphi^{-1}(\tau))}{\sqrt{\dot \varphi^{-1}(\tau)}}\cosh\theta(\varphi^{-1}(\tau)),\,\frac{r(\varphi^{-1}(\tau))}{\sqrt{\dot \varphi^{-1}(\tau)}}\sinh\theta(\varphi^{-1}(\tau))\right)\, .
\end{equation}
It commutes with the Lorentz transformations (\ref{Lorentz})

Now the integral
\begin{equation}
    \frac{1}{\rho^{2}}=\frac{1}{T}\int_0^T\frac{d\tau}{(\xi^0(\tau))^2-(\xi^1(\tau))^2}
\end{equation}
is invariant under the transformations (\ref{pseudo act}).
On every orbit of the action of the group $Diff_{+}^1([0,T])$, we choose an element of the form
\begin{equation}\label{one element pseudo}
    \tilde \xi_{\rho,\psi}=(\rho\cosh \psi(\tau),\rho\sinh \psi(\tau))\, ,
\end{equation}
where $\psi\in C([0,T])$. Any element $\xi\in C([0,T],\mathrm{Cone})$ is obtained from the element of the form (\ref{one element pseudo}) using the transformation (\ref{pseudo act}).
\begin{equation}
    \xi(\tau)=(\varphi\tilde \xi_{\rho,\psi})(\tau)=\left(\frac{\rho}{\sqrt{\dot\varphi^{-1}(\tau)}}\cosh\psi(\varphi^{-1}(\tau))\, ,\, \frac{\rho}{\sqrt{\dot \varphi^{-1}(\tau)}}\sinh\psi(\varphi^{-1}(\tau))\right)
\end{equation}

Thus, every element $\xi \in C([0,T],\mathrm{Cone})$ corresponds to a triple $(\rho,\psi,\varphi)$, where $\rho>0,\, \psi(\tau)\in C([0,T])$,\, $\varphi\in Diff^1_{+}([0,T])$
\begin{equation}
    \xi\longleftrightarrow(\rho,\psi,\varphi)\, .
\end{equation}
\begin{equation}\label{decompcone}
    \frac{1}{\rho^2}=\frac{1}{T}\int_0^T\frac{d\tau_1}{(\xi^0(\tau_1))^2-(\xi^1(\tau_1))^2}\, ,\qquad \varphi^{-1}(\tau)=\frac{\int_0^{\tau}\frac{d\tau_1}{(\xi^0(\tau_1))^2-(\xi^1(\tau_1))^2}}{\frac{1}{T}\int_0^T\frac{d\tau_1}{(\xi^0(\tau_1))^2-(\xi^1(\tau_1))^2}}\, ,
\end{equation}
\begin{equation*}
    \psi(s)=\mathrm{arctanh}\left(\frac{\xi^1(\varphi(s))}{\xi^0(\varphi(s))}\right)\, .
\end{equation*}
It is easy to verify the validity of the relation
\begin{equation}\label{Lorentz_inv}
    L_{\gamma}(\rho,\psi,\varphi)\longleftrightarrow(\rho,\psi+\gamma,\varphi)\, .
\end{equation}

This correspondence allows us, by analogy with the Wiener measure on $C([0,T],\mathbb{R}^2)$, to define a measure on $C([0,T],\mathrm{Cone})$ using a decomposition over the orbits of the group of diffeomorphisms
\begin{equation}\label{wtilde}
     \tilde w_{\sigma}(d\xi)=\rho\exp\left(-\frac{\sigma^2}{4\rho^2}\right)\dot \varphi(0)\dot \varphi(T) \mu_{\frac{2\sigma}{\rho}}(d\varphi)w_{\frac{\sigma}{\rho}}(d\psi)d\rho
\end{equation}
Now from the (\ref{Lorentz_inv}), the invariance of the measure $\tilde w_{\sigma}$ (\ref{wtilde}) under the transformations (\ref{Lorentz}) is obvious.

In Cartesian coordinates $\xi^0(\tau)$, $\xi^1(\tau)$ the measure $\tilde w_{\sigma}$ has the form (see Appendix B)
\begin{equation}\label{tildewxi}
    \tilde w_{\sigma}(d\xi)=\exp\left(-\frac{1}{2\sigma^2}\int_0^T \langle \dot \xi,\dot \xi\rangle \,\,d\tau\right)d\xi\,  ,
\end{equation}
where
\begin{equation}\label{Conemetric}
    \langle \dot \xi,\dot \xi\rangle\equiv(\dot \xi^0)^2-(\dot \xi^1)^2+2\frac{(\xi^0\dot \xi^1-\xi^1\dot \xi^0)^2}{(\xi^0)^2-(\xi^1)^2}\, .
\end{equation}

For the constructed measure $\tilde w_{\sigma}$, an equality holds, analogous to (\ref{1ddecomp}), (\ref{2ddecomp}). We split the random process $\xi\in C([0,T],\mathbb{R}^2)$ into two subprocesses: let $\xi_1(\tau)$ coincide with $\xi(\tau)$ on the interval $[0,t^{*}]$, and $\xi_2(\tau)$ coincide with $\xi(\tau)$ on the interval $[t^{*},T]$. To each of the subprocesses we associate the set of decomposition coordinates
\begin{equation}
    \xi_1\longleftrightarrow(\rho_1,\psi_1,\varphi_1)\, ,\qquad \xi_2\longleftrightarrow(\rho_2,\psi_2,\varphi_2)\, .
\end{equation}
The coordinates $(\rho_1,\psi_1,\varphi_1)$ and $(\rho_2,\psi_2,\varphi_2)$ are uniquely expressed in terms of the coordinates $(\rho,\psi,\varphi)$ of the original process
\begin{equation}\label{Omegacone}
    (\rho,\psi,\varphi)=\Omega^{\mathrm{Cone}}(\rho_1,\psi_1,\varphi_1,\rho_2,\psi_2,\varphi_2)
\end{equation}
(the explicit form of $\Omega^{\mathrm{Cone}}$ is given in Appendix A). Let $\Phi(\rho,\psi,\varphi)$ be an arbitrary functional on $\mathbb{R}^{+}\times C([0,T])\times Diff_{+}^{1}([0,T])$. We construct a functional $\tilde \Phi(\rho_1,\psi_1,\varphi_1,\rho_2,\psi_2,\varphi_2)$ as follows
\begin{equation}
    \tilde \Phi(\rho_1,\psi_1,\varphi_1,\rho_2,\psi_2,\varphi_2)=\Phi(\Omega^{\mathrm{Cone}}(\rho_1,\psi_1,\varphi_1,\rho_2,\psi_2,\varphi_2))\,.
\end{equation}
Let $\mu^1_{\sigma}$ be the quasi-invariant measure on $Diff^1_{+}([0,t^{*}])$, $\mu^2_{\sigma}$ be the quasi-invariant measure on $Diff^1_{+}([t^{*},T])$, $w^{1}_{\sigma}$ be the Wiener measure on $C([0,t^{*}])$, $w^{2}_{\sigma}$ be the Wiener measure on $C([t^{*},T])$. Then the following equality is valid

\begin{equation}\label{casuality_cone}
    \int \Phi(\rho,\varphi,\psi)\,\,\rho \exp\left(-\frac{\sigma^2}{4\rho^2}\right)\,\dot \varphi(0)\dot \varphi(T)\, \mu_{\frac{2\sigma}{\rho}}(d\varphi)\,w_{\frac{\sigma}{\rho}}(d\psi)\,d\rho=
\end{equation}
\begin{equation*}
    =\int \tilde \Phi(\rho_1,\varphi_1,\psi_1,\rho_2,\varphi_2,\psi_2)\frac{\delta\left(\rho_1\sqrt{\dot \varphi_1(t^{*})}-\rho_2\sqrt{\dot \varphi_2(t^{*})}\right)}{\rho_1\sqrt{\dot \varphi_1(t^{*})}}\delta\left(\psi_2(t^{*})-\psi_1(t^*)\right)
\end{equation*}
\begin{equation*}
    \rho_1\rho_2\,\exp\left(-\frac{\sigma^2}{4\rho_1^2}-\frac{\sigma^2}{4\rho_2^2}\right)\,\dot \varphi_1(0)\dot \varphi_1(t^{*})\dot \varphi_2(t^{*})\dot \varphi_2(T)\,
\end{equation*}
\begin{equation*}
    \mu^1_{\frac{2\sigma}{\rho_1}}(d\varphi_1)\,\mu^{2}_{\frac{2\sigma}{\rho_2}}(d\varphi_2)\,w^1_{\frac{\sigma}{\rho_1}}(d\psi_1)\,w^2_{\frac{\sigma}{\rho_2}}(d\psi_2)\,d\rho_1\,d\rho_2\, .
\end{equation*}

\section{Random process on the infinite-sheeted covering of the plane}

\subsection{Metric on the cone and its properties}
The quadratic form (\ref{Conemetric}) defines a metric on the cone. Let us describe the properties of this metric. The line element has the form\footnote{Here we denote coordinates on $\mathrm{Cone}$ as $(x^0,x^1)$. The letter $\xi$ is used for a random process on $\mathrm{Cone}$}
\begin{equation}\label{Conemet1}
    ds^2\equiv\langle dx,dx\rangle= (dx^0)^2-(dx^1)^2+2\frac{(x^0dx^1-x^1dx^0)^2}{(x^0)^2-(x^1)^2}\,=
\end{equation}
\begin{equation*}
    =\frac{(x^0)^2+(x^1)^2}{(x^0)^2-(x^1)^2}\left((dx^0)^2+(dx^1)^2\right)-4\frac{x^0x^1}{(x^0)^2-(x^1)^2}dx^0 dx^1\, =
\end{equation*}
\begin{equation*}
    =x^{+}x^{-}\left(\left(\frac{dx^{+}}{x^{+}}\right)^2+\left(\frac{dx^{-}}{x^{-}}\right)^2\right)=
\end{equation*}
\begin{equation*}
    =e^{l^{+}+l^{-}}((dl^{+})^2+(dl^{-})^2)\, ,
\end{equation*}
where
\begin{equation}
    x^{\pm}=\frac{x^0\pm x^1}{\sqrt{2}}
\end{equation}
and 
\begin{equation}
    l^{\pm}=\ln x^{\pm}\, .
\end{equation}

In the Minkowskian coordinates $(r,\theta)$
\begin{equation}\label{x0x1hyp}
    x^0=r\cosh\theta\,, \qquad x^1=r\sinh\theta\, .
\end{equation}
the metric (\ref{Conemet1}) looks like the Euclidean metric on $\mathbb{R}^2$
\begin{equation}\label{polarmet}
    ds^2=dr^2+r^2d\theta^2\, .
\end{equation}
Note that in eqs (\ref{x0x1hyp}), (\ref{polarmet}), the "angular" coordinate $\theta$ is real value ($\theta\in\mathbb{R}$).

\subsection{Isometric mapping of the cone onto the infinite-sheeted covering of the plane}

The space $\mathrm{Cone}$ with the metric (\ref{Conemet1}) defined on it admits a bijective isometric mapping onto the infinite-sheeted covering of the plane. Namely, consider a set of planes
\begin{equation}
    \Pi_n=\{(r,\theta)| r>0,\theta\in [0,2\pi)\}\, ,
\end{equation}
where each point of the plane $\Pi_n$ is given by a pair of its polar coordinates: the distance from the origin $r$ and the polar angle $\theta$. On each of the planes $\Pi_n$, the Euclidean metric of the form (\ref{polarmet}) is defined. On all planes $\Pi_n$, we make a cut along the direction $\theta=0$ and glue the planes along these cuts so that a point on plane $\Pi_n$, after making a full counterclockwise revolution around the origin, ends up on plane $\Pi_{n+1}$. In other words, we set
\begin{equation}
    \lim_{\theta\to2\pi} (r,\theta)_n=(r,0)_{n+1}\, .
\end{equation}
The collection of planes $\Pi_n$ glued together in this manner will be denoted as $\mathrm{Cover}$ and called the infinite-sheeted covering of the plane. Each point of $\mathrm{Cover}$ can be specified by a triple $(r,\theta,n)$, where $r,\theta$ are the polar coordinates on one of the planes $\Pi_n$, with $\theta\in[0,2\pi)$, and $n$ is the plane number. Alternatively, one can specify a point by a pair $(r,\theta)$, where $\theta\in \mathbb{R}$ and adopt the convention that if $2\pi n\le \theta<2\pi(n+1)$, then the point belongs to $\Pi_n$, and the polar angle on $\Pi_n$ is $\theta-2\pi n$.

Thus, each point of $\mathrm{Cover}$ is given by two coordinates $r\in \mathbb{R}^{+}$, $\theta\in \mathbb{R}$. The metric on $\mathrm{Cover}$ is inherited from each of the planes $\Pi_{n}$ and therefore has the form (\ref{polarmet}). It is easy to see that there exists a one-to-one correspondence between $\mathrm{Cone}$ and $\mathrm{Cover}$, which is given by a coordinate-wise matching of points
\begin{equation}\label{conecov}
    (r,\theta)\in \mathrm{Cone}\longleftrightarrow(r,\theta)\in\mathrm{Cover}\, .
\end{equation}
This correspondence is isometric since the metric on $\mathrm{Cone}$ (\ref{polarmet}) coincides with the metric on $\mathrm{Cover}$ (\ref{polarmet}). Under this mapping, each plane $\Pi_n$ corresponds to a sector on $\mathrm{Cone}$ (the sector enclosed between the red dashed lines in Fig. 1).

\begin{figure}[H]
    \centering
    \includegraphics[width=0.5\linewidth]{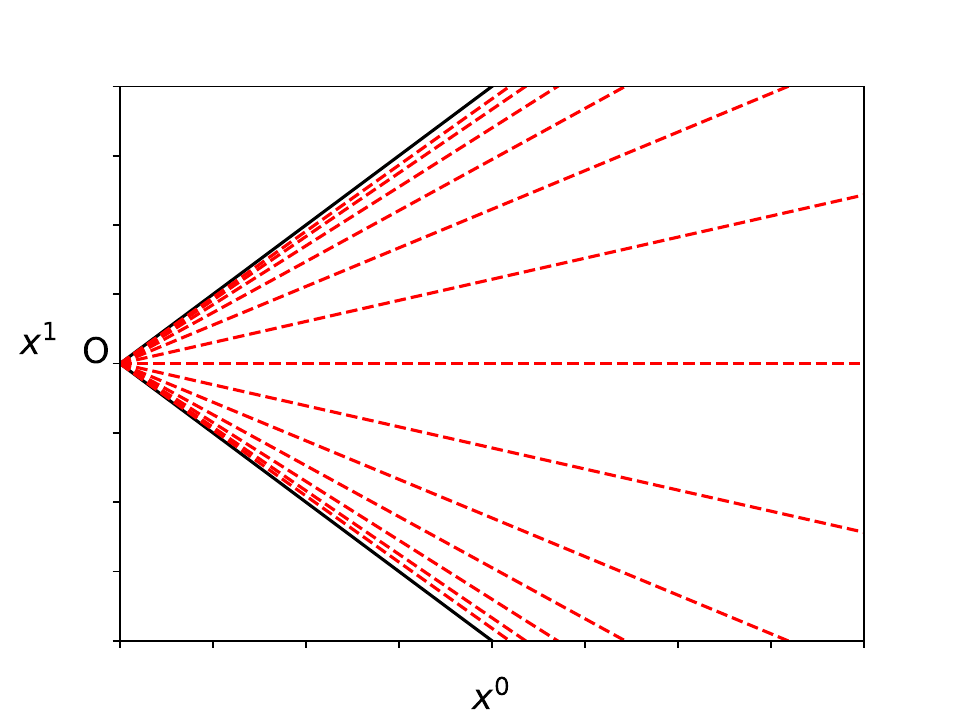}
    \caption{}
    \label{fig:placeholder}
\end{figure}

\newpage

\subsection{Geodesics on the cone}
Consider a functional integral with respect to the measure $\tilde w_{\sigma}$ of the following form
\begin{equation}\label{KAB}
    K(A,B)=\int \delta(\xi(0)-A)\delta(\xi(T)-B)\tilde w_{\sigma}(d\xi)\, ,
\end{equation}
where $A$, $B$ are two fixed points on the cone. Such integrals are often encountered in quantum field theory and describe the propagation of a particle from point $A$ to point $B$. As $\sigma\to 0$ (which is an analogue of the semiclassical approximation), the main contribution to integrals of the form (\ref{KAB}) is given, as is known, by geodesics of the metric (\ref{Conemet1}). Therefore, consider the following problem. Let two points $A$ with pseudo-polar coordinates $(r_A,\theta_A)$ and $B$ with pseudo-polar coordinates $(r_B,\theta_B)$ be given on $\mathrm{Cone}$. What does the geodesic passing through these two points look like? It turns out that the geodesics have qualitatively different forms depending on the angular distance $|\theta_A-\theta_B|$. Let us consider three cases.

\vspace{0.5cm}

{\bf Case 1: $|\theta_A-\theta_B|\le \pi$}\\
Since the metric (\ref{Conemet1}), written in pseudo-polar coordinates (\ref{x0x1hyp}), has the form (\ref{polarmet}), which coincides with the metric of the Euclidean plane, the equation of geodesics in coordinates $r,\theta$ coincides with the equation of a straight line on the plane in polar coordinates
\begin{equation}\label{geodpl1}
    r\cos(\theta-\theta_0)=r_0\, .
\end{equation}
In this equation, the coordinate $\theta$ takes values
\begin{equation}
    \theta\in \left[-\frac{\pi}{2}+\theta_0,\frac{\pi}{2}+\theta_0\right]\, .
\end{equation}
It follows that two points $A,B\in \mathrm{Cone}$ can be connected by a smooth geodesic of the form (\ref{geodpl1}) only if the angular distance between the points does not exceed $\pi$: $|\theta_A-\theta_B|\le\pi$.
Note that under the mapping of $\mathrm{Cone}$ onto $\mathrm{Cov}_{\infty}$ constructed above, both points $A$, $B$ fall on the same sheet of the covering, and the geodesic of the form (\ref{geodpl1}) maps to a straight line (Fig. 2).

\begin{figure}[H]
    \centering
    \begin{subfigure}{0.45\textwidth}
        \centering
        \includegraphics[width=\linewidth]{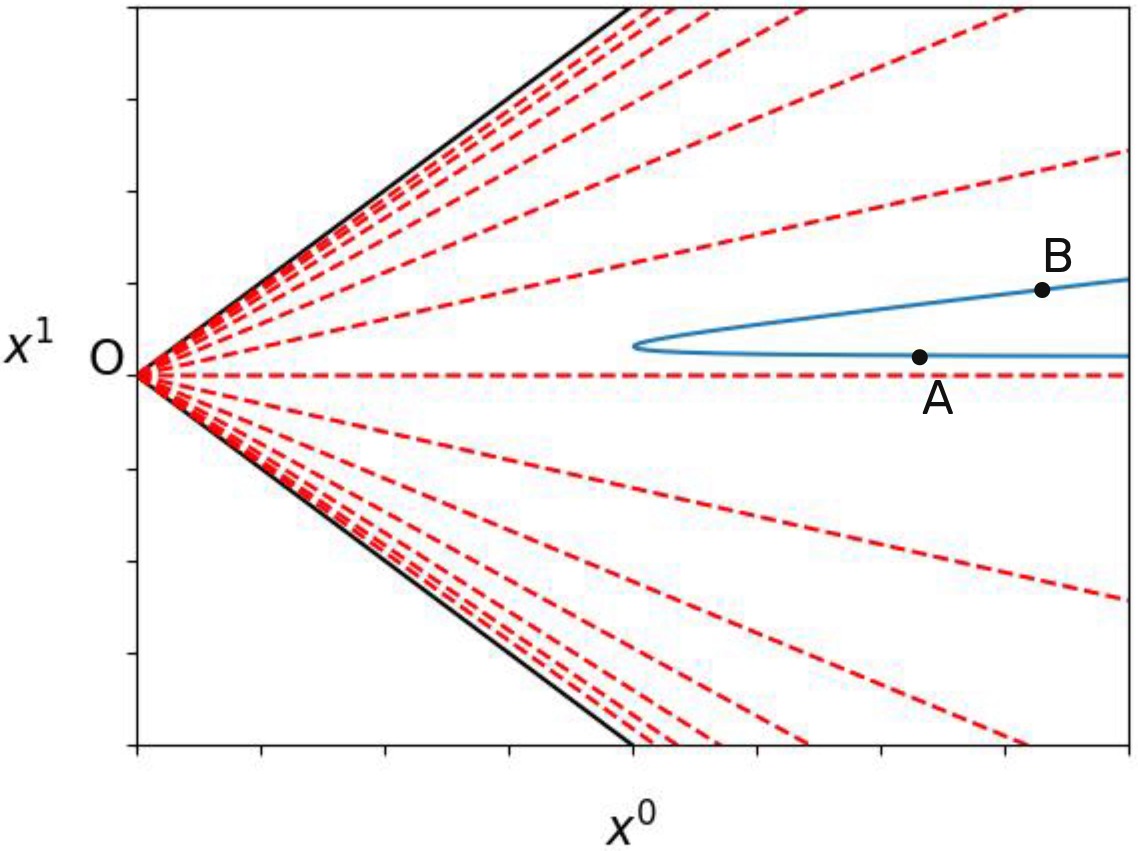}
        \caption{}
    \end{subfigure}
    \hfill
    \begin{subfigure}{0.45\textwidth}
        \centering
        \includegraphics[width=\linewidth]{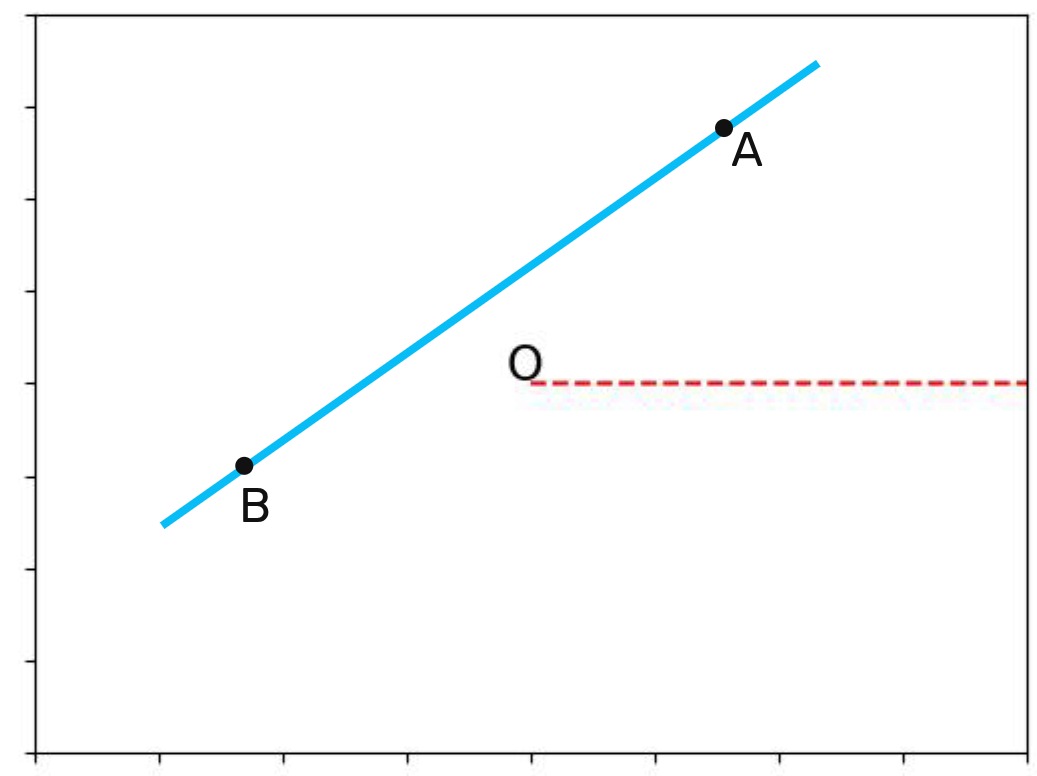}
        \caption{}
    \end{subfigure}
    \hfill
    \caption{Geodesic passing through points $A$ and $B$ in the case where the angular distance between them is less than $\pi$. Figure (a) shows the geodesic on the cone, figure (b) shows its image under the mapping onto the infinite-sheeted covering of the plane.}
    \label{fig:three_images_case1}
\end{figure}

\newpage

{\bf Case 2: $\pi<|\theta_A-\theta_B|<2\pi$}\\
  It is convenient to consider simultaneously points $A,B\in \mathrm{Cone}$ and their images on $\mathrm{Cover}$. When $|\theta_A-\theta_B|>\pi$, both points $A$, $B$ fall on the same sheet of the covering. On this sheet, one can easily construct a geodesic. Due to the presence of a cut on the covering sheet, the straight line $AB$ is not a geodesic because it would intersect the cut. Any segment that does not intersect the cut has the smallest length among all curves connecting its endpoints. Therefore, the curve of minimal length connecting points $A$ and $B$ is a broken line consisting of two segments connecting points $A$ and $B$ to the origin $O$ (Fig. 3).

\begin{figure}[H]
    \centering
    \begin{subfigure}{0.45\textwidth}
        \centering
        \includegraphics[width=\linewidth]{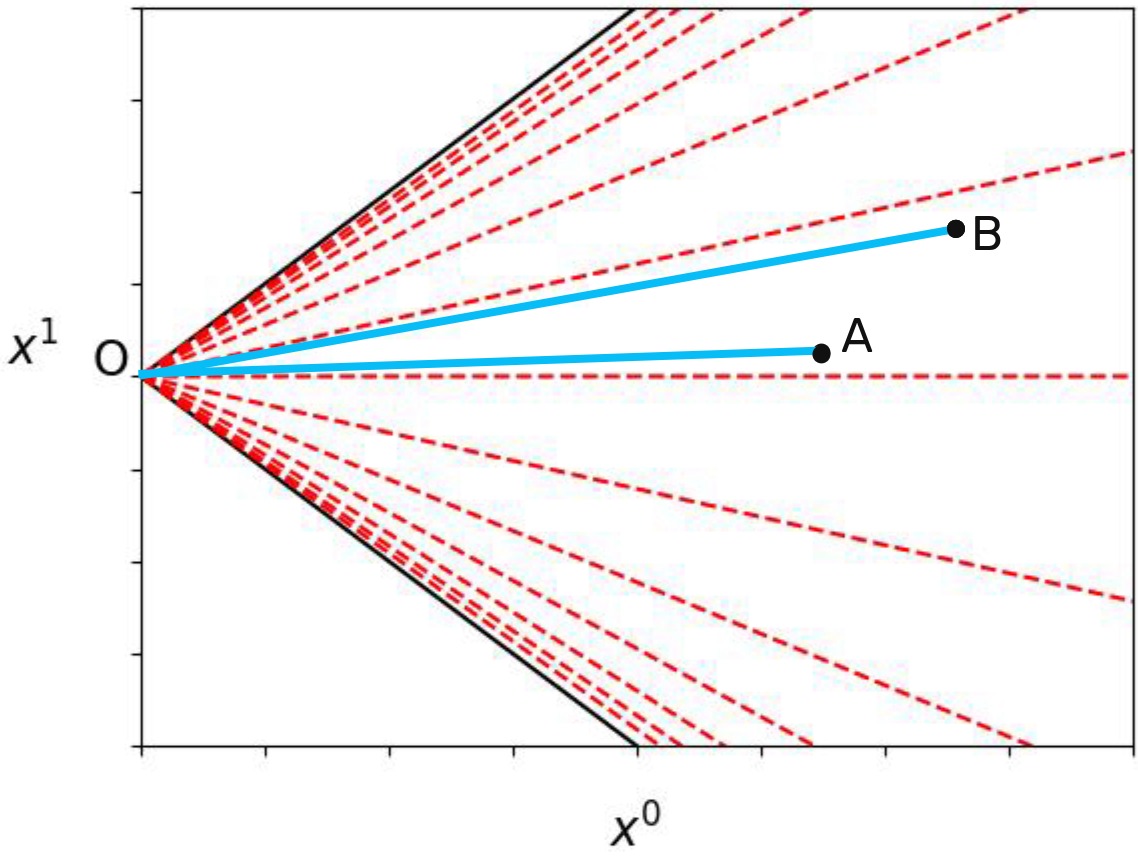}
        \caption{}
    \end{subfigure}
    \hfill
    \begin{subfigure}{0.45\textwidth}
        \centering
        \includegraphics[width=\linewidth]{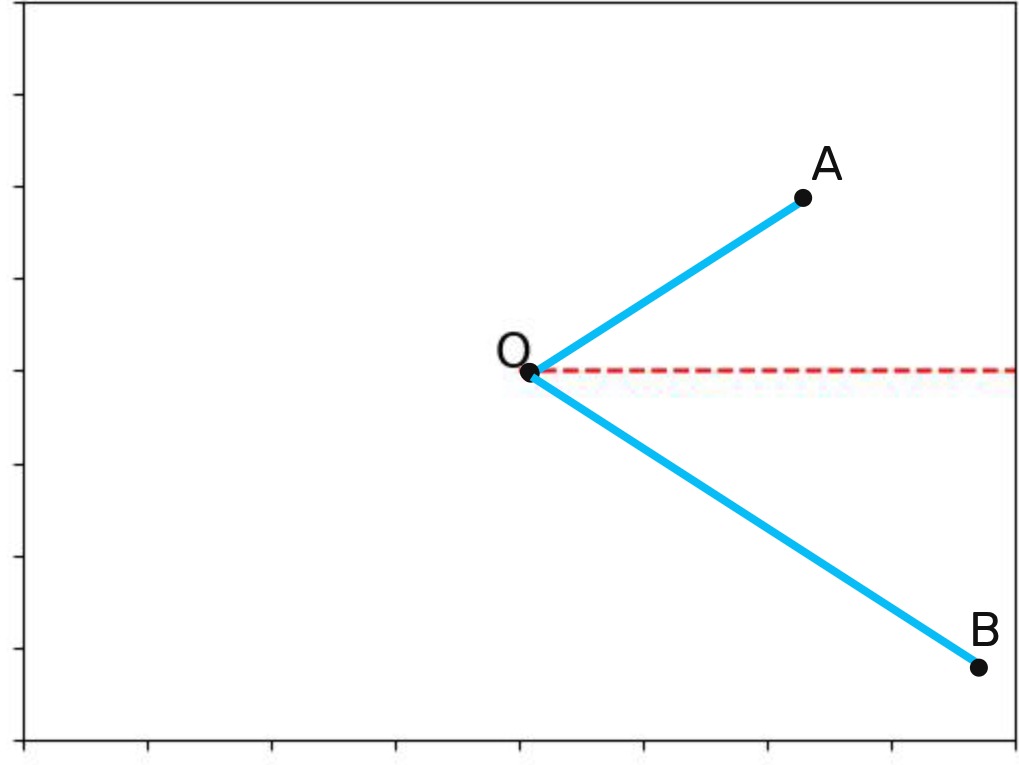}
        \caption{}
    \end{subfigure}
    \hfill
    \caption{Geodesic passing through points $A$ and $B$ in the case where the angular distance between them is greater than $\pi$ but less than $2\pi$. Figure (a) shows the geodesic on the cone, figure (b) shows its image under the mapping onto the infinite-sheeted covering of the plane. The images of points $A$ and $B$ lie on the same sheet of the covering.}
    \label{fig:three_images_case2}
\end{figure}

\newpage

{\bf Case 3: $|\theta_A-\theta_B|>2\pi$} \\
  In this case, points $A,B$ fall on different sheets of the covering, and the geodesic, as in the previous case, is a broken line consisting of two segments $OA$ and $OB$. Indeed, any other curve connecting points $A$ and $B$ (Fig. 4 (a)) under the mapping onto $\mathrm{Cover}$ will be represented as two arcs on different sheets of the covering, passing around the branch point $O$ (Fig. 4 (b) (c)). Obviously, $AO+OB<AM+MB$.

\begin{figure}[H]
    \centering
    \begin{subfigure}{0.45\textwidth}
        \centering
        \includegraphics[width=\linewidth]{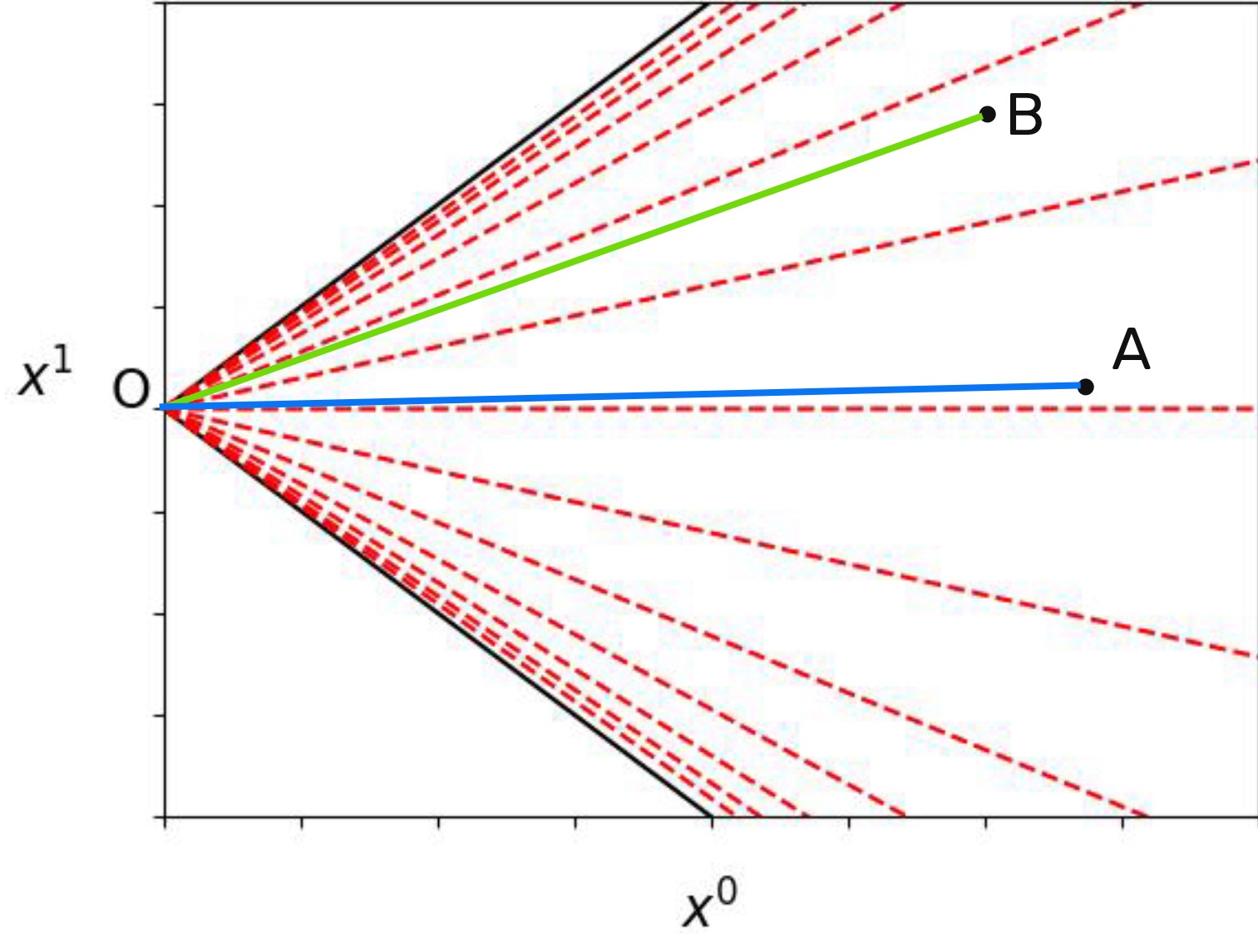}
        \caption{}
    \end{subfigure}
    \hfill
    \begin{subfigure}{0.45\textwidth}
        \centering
        \includegraphics[width=\linewidth]{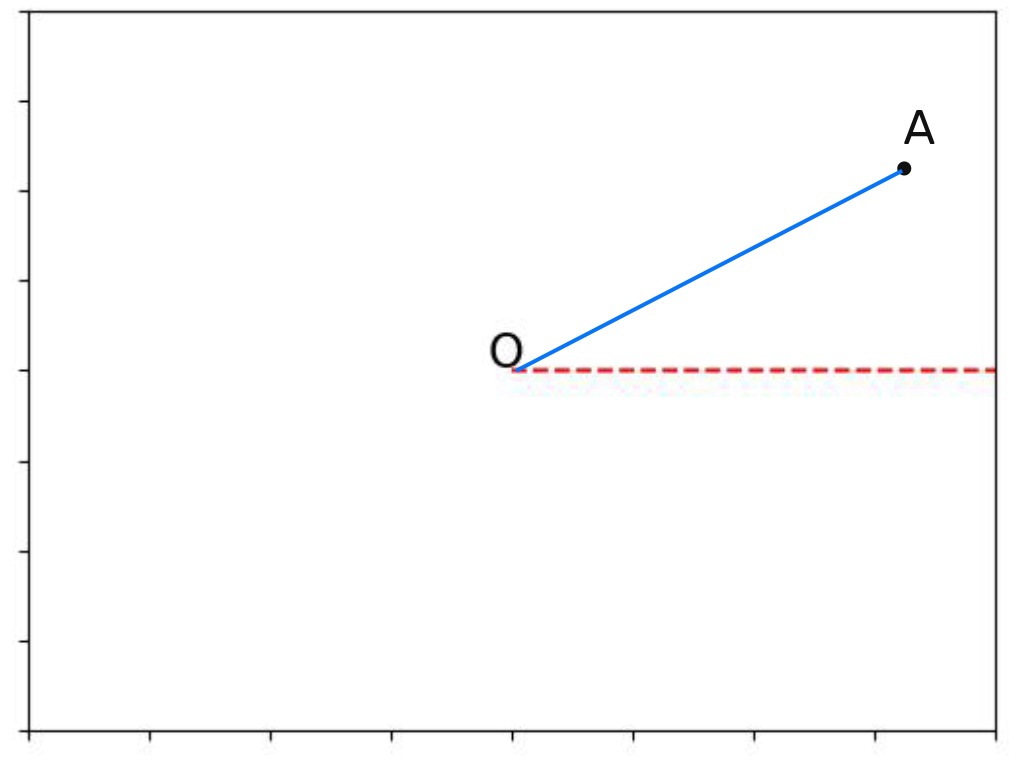}
        \caption{}
    \end{subfigure}
    \hfill
    \begin{subfigure}{0.45\textwidth}
        \centering
        \includegraphics[width=\linewidth]{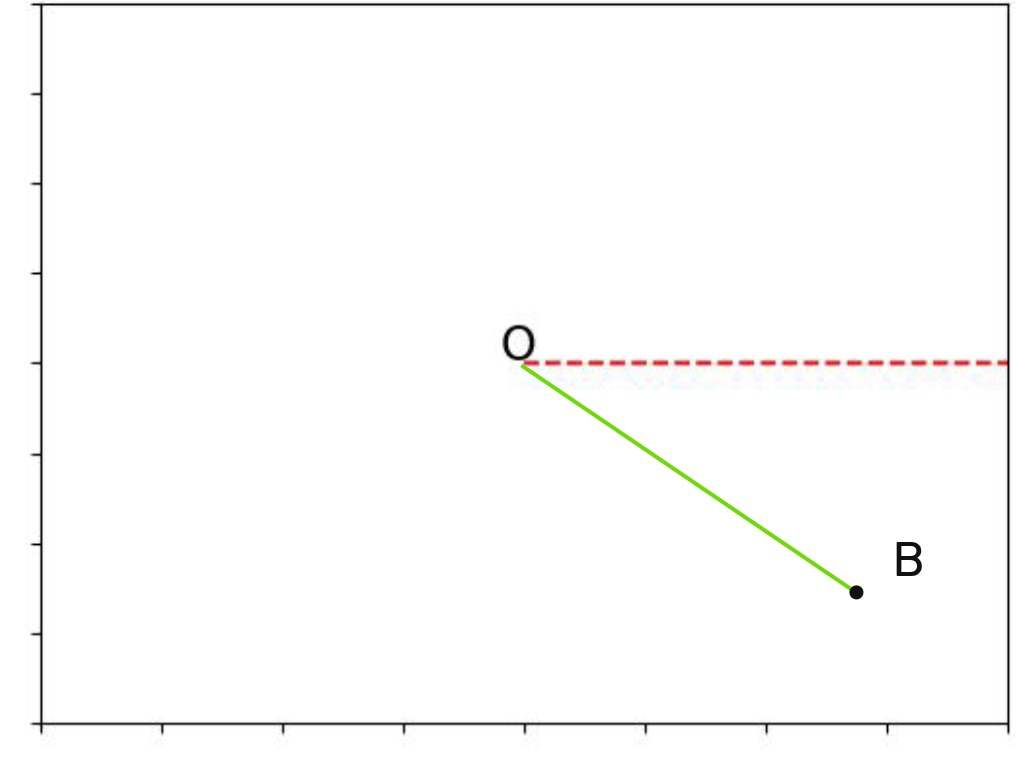}
        \caption{}
    \end{subfigure}
    \caption{Geodesic passing through points $A$ and $B$ in the case where the angular distance between them is greater than $2\pi$. Figure (a) shows the geodesic on the cone, figures (b) and (c) show its image under the mapping onto the infinite-sheeted covering of the plane. The images of points $A$ and $B$ lie on different sheets of the covering.}
    \label{fig:three_images_case3}
\end{figure}

\begin{figure}[H]
    \centering
    \begin{subfigure}{0.45\textwidth}
        \centering
        \includegraphics[width=\linewidth]{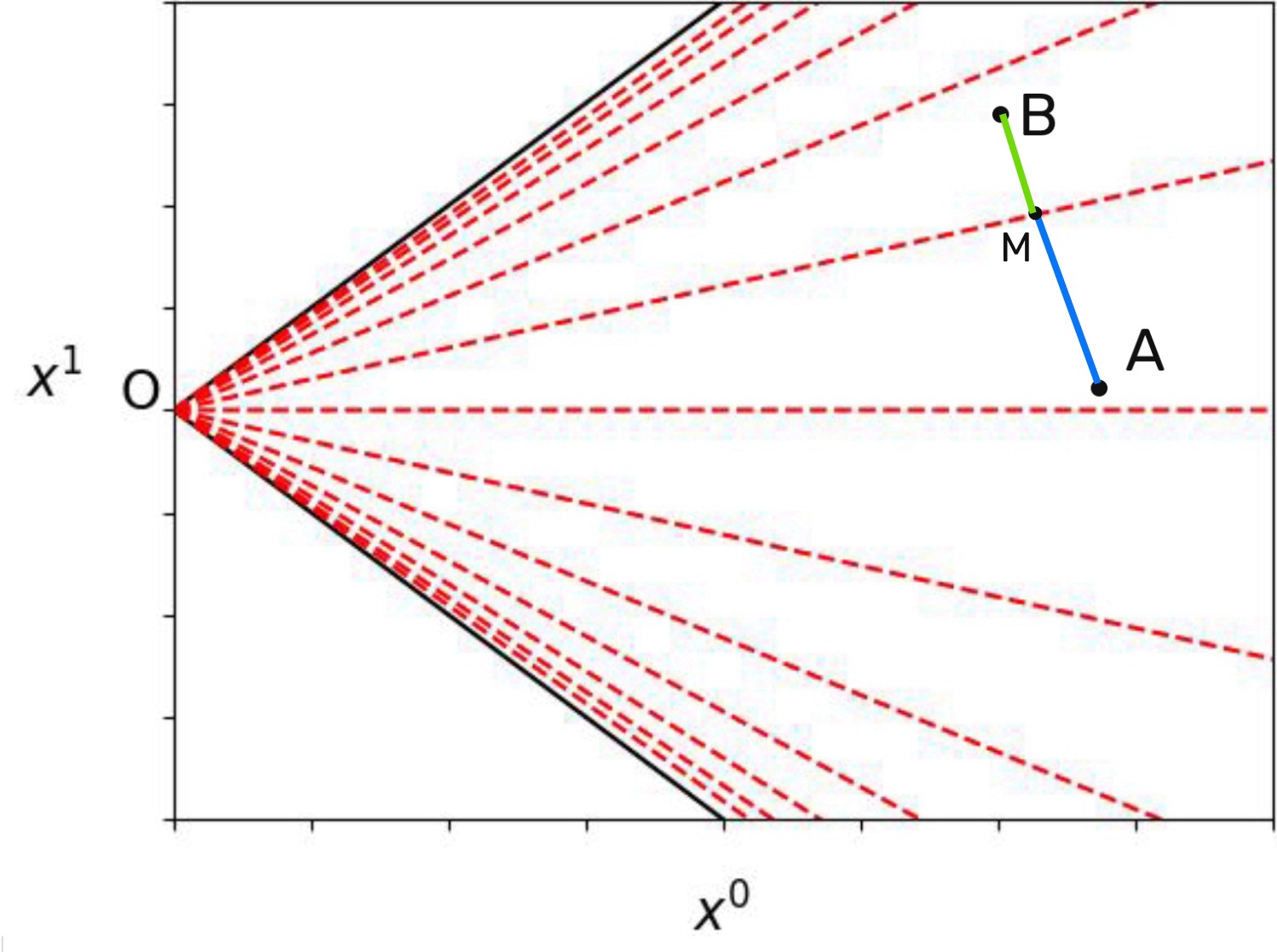}
        \caption{}
    \end{subfigure}
    \hfill
    \begin{subfigure}{0.43\textwidth}
        \centering
        \includegraphics[width=\linewidth]{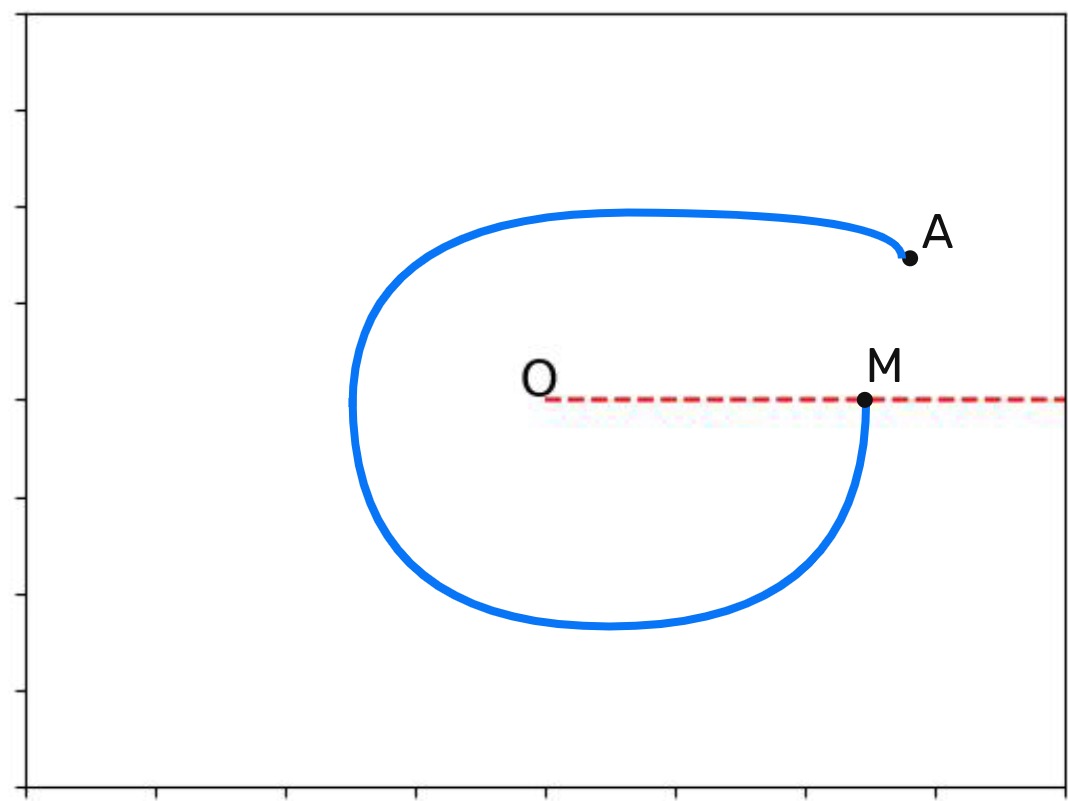}
        \caption{}
    \end{subfigure}
    \hfill
    \begin{subfigure}{0.43\textwidth}
        \centering
        \includegraphics[width=\linewidth]{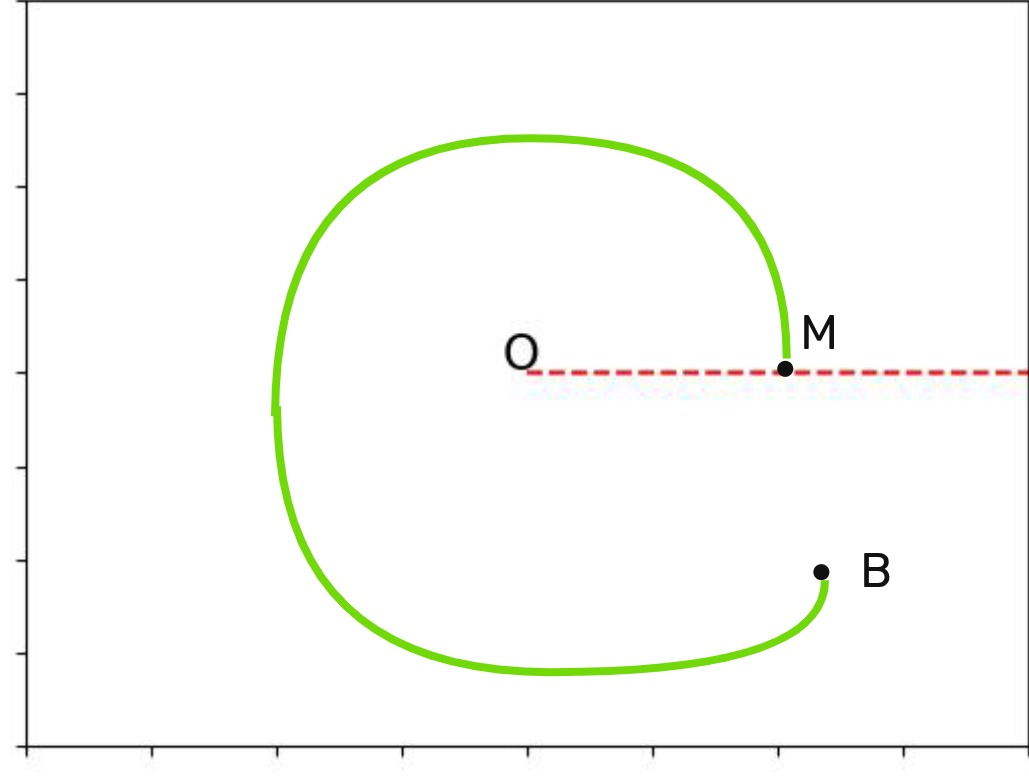}
        \caption{}
    \end{subfigure}
    \caption{Example of a curve connecting points $A$ and $B$ whose angular distance exceeds $2\pi$. Figure (a) shows the curve on the cone, figures (b) and (c) show its image on the infinite-sheeted covering. The images of points $A$ and $B$ lie on different sheets of the covering.}
    \label{fig:three_images_other_curve}
\end{figure}

The correspondence (\ref{conecov}) allows us to consider the constructed measure (\ref{wtilde}) as a measure on $C([0,T],\mathrm{Cover})$. Thus, we have obtained a tool for describing the motion of a quantum particle on the infinite-sheeted covering of the plane. When calculating integrals of the form (\ref{KAB}), which describe the propagation of a particle from point $A$ to point $B$ lying on the same sheet of the covering, trajectories that cross the cut and go out to other sheets of the covering will also contribute to the integral (Fig. 6)\footnote{It looks like the 3-d season of the famous film by David Lynch and explains the title of the present paper.} .

\begin{figure}[H]
    \centering
    \begin{subfigure}{0.45\textwidth}
        \centering
        \includegraphics[width=\linewidth]{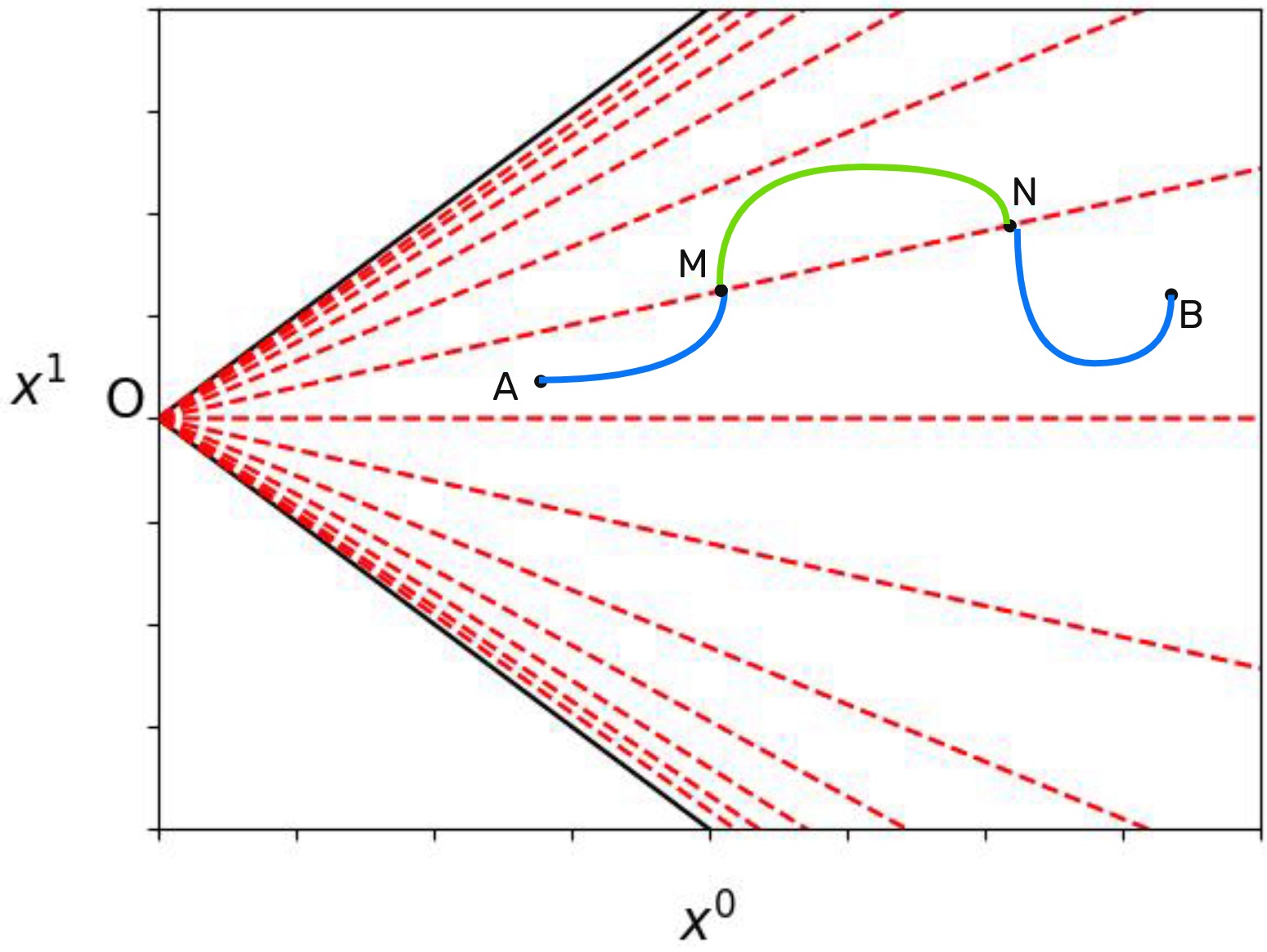}
        \caption{}
    \end{subfigure}
    \hfill
    \begin{subfigure}{0.43\textwidth}
        \centering
        \includegraphics[width=\linewidth]{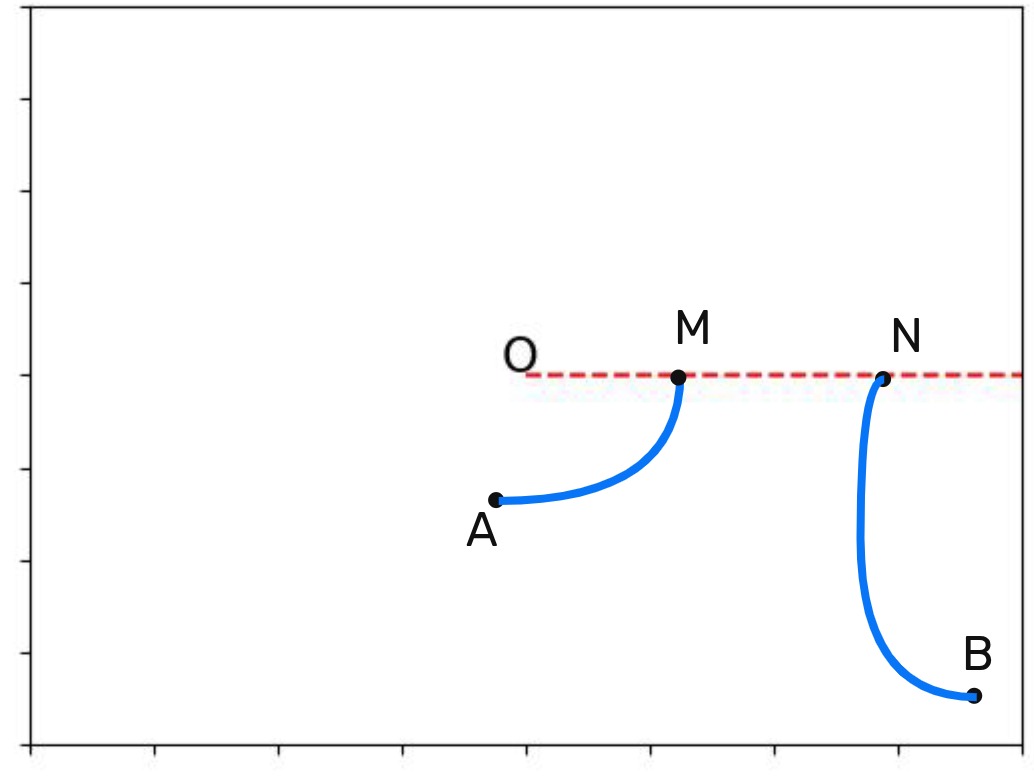}
        \caption{}
    \end{subfigure}
    \hfill
    \begin{subfigure}{0.43\textwidth}
        \centering
        \includegraphics[width=\linewidth]{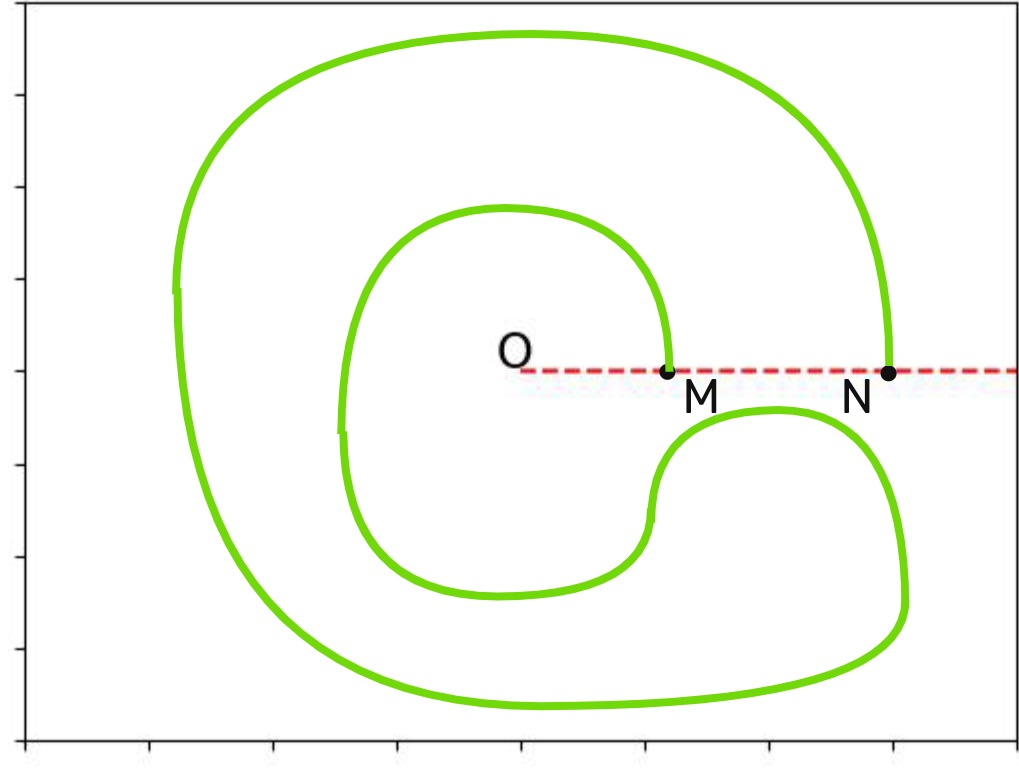}
        \caption{}
    \end{subfigure}
    \caption{Example of a trajectory connecting two points on the same sheet of the covering, but lying on different sheets. The trajectory on the cone is shown in figure (a), its image on the covering is shown in the figures (b), (c).}
    \label{fig:three_images_trajectory}
\end{figure}

\section{Conclusion}
In the present paper, the measure on trajectories lying inside the cone has been constructed
\begin{equation}
    \tilde w_{\sigma}(d\xi)=\rho\,\exp\left(-\frac{\sigma^2}{4\rho^2}\right)\,\dot \varphi(0)\dot \varphi(T)\, \mu_{\frac{2\sigma}{\rho}}(d\varphi)\,w_{\frac{\sigma}{\rho}}(d\psi)\,d\rho=\exp\left(-\frac{1}{2\sigma^2}\int_0^Td\tau\,\, \langle \dot \xi, \dot \xi \rangle\right)\,d\xi\, ,
\end{equation}
where $\langle\dot \xi,\dot \xi \rangle$ is given by formula (\ref{Conemetric}). This measure is invariant under Lorentz transformations (\ref{Lorentz}) and quasi-invariance under the action of the group of diffeomorphisms (\ref{pseudo act}). The constructed measure also possesses the causality property (\ref{casuality_cone}), which allows one to average a certain functional $F(\xi)$ over the values of the random process on the interval $[t^{*},T]$ independently of the values of this process on the interval $[0,t^{*})$.

It is shown that the space $\mathrm{Cone}$ with the metric (\ref{Conemetric}) defined on it admits a one-to-one isometric mapping onto the infinite-sheeted covering of the plane $\mathrm{Cover}$. Geodesics on these spaces are described; trajectories in their vicinity give the main contribution to the path integral as $\sigma\to 0$.

The constructed measure will make it possible to compute the propagator in Euclidean quantum field theory \cite{Gibbons}--\cite{Witten ph} in a space with a nontrivial metric (\ref{Conemetric}). This computation will be presented in the subsequent paper. Let us outline the main idea here. The propagator satisfies an equation of the form \cite{Buhbind}
\begin{equation}\label{propeq}
    (\Delta-m^2)G(x,y)=\delta(x-y)\, ,
\end{equation}
where $\Delta$ is the Laplace-Beltrami operator constructed from the corresponding metric $g_{\mu\nu}$. The propagator $G(x,y)$ is uniquely expressed through the solution of a heat-type equation \cite{Kaku},\cite{Padm}
\begin{equation}
G(x,y)=\int_{\mathbb{R}}K(x,y,\tau)e^{-m^2\tau}d\tau\, ,
\end{equation}
\begin{equation}\label{Heat}
    \frac{\partial K(x,y,\tau)}{\partial \tau}=\Delta K+\delta(x-y)\, .
\end{equation}
 $K(x,y,\tau)$ can be represented as a path integral of the form\footnote{Note that the integration is carried out not over the space of fields $\phi(x)$ depending on $d$ variables, but over the space of particle trajectories $\xi(\tau)$ depending on a single variable. The connection between integration over the space of fields and corresponding particles was discussed in detail in \cite{Padm}.}
\begin{equation}\label{KPI}
    K(x,y,\tau)=\int \delta(\xi(0)-x)\delta(\xi(\tau)-y)\nu(d\xi)\, ,
\end{equation}
where $\nu$ is an integration measure defined on trajectories in the space with metric $g_{\mu\nu}$ (in the case of Euclidean space it coincides with the Wiener measure).

The obtained results may be used in the problem of thermal radiation from black holes. The radial part of the Schwarzschild metric near the black hole horizon can be represented in the form \cite{Gibbons}, \cite{Boulware}
\begin{equation}
    ds^2=-r^2dt^2+dr^2
\end{equation}
(the Rindler metric). In the subsequent paper using the decomposition of the measures (\ref{win2d}), (\ref{wtilde}) we define and calculate path integrals in such theories.

\vspace{1cm}
The research has been performed in the framework of the state assignment of M.V. Lomonosov Moscow State University. The work of V. Chistiakov is supported by "Basis" foundation.

\section{Appendix A}
We present the explicit form of the mappings introduced in eqs. (\ref{OmegaR}), (\ref{OmegaR2}), (\ref{Omegacone}). The mapping $(\rho_1,\varphi_1,\rho_2,\varphi_2)=\Omega^{-1}(\rho,\varphi)$ has the form
\begin{equation}\label{rho12}
    \rho_1=\rho\sqrt{\frac{t^{*}}{\varphi^{-1}(t^{*})}}\, ,\qquad \rho_2=\rho\sqrt{\frac{T-t^{*}}{T-\varphi^{-1}(t^{*})}}\, .
\end{equation}
\begin{equation}\label{phi12}
    \varphi^{-1}_1(\tau)=\frac{t^{*}}{\varphi^{-1}(t^{*})}\varphi^{-1}(\tau)\, ,\qquad \varphi_2^{-1}(\tau)=\frac{T-t^{*}}{T-\varphi^{-1}(t^{*})}(\varphi^{-1}(\tau)-\varphi^{-1}(t^{*}))+t^{*}\, .
\end{equation}
Conversely, given $(\rho_1,\varphi_1,\rho_2,\varphi_2)$, one can uniquely reconstruct the decomposition coordinates $(\rho,\varphi)$
\begin{equation}\label{rhophi}
    \frac{1}{\rho^2}=\frac{t^{*}}{T}\frac{1}{\rho_1^2}+\frac{T-t^{*}}{T}\frac{1}{\rho_2^2}\, ,\qquad \varphi^{-1}(\tau)=\left\{\begin{array}{cc}
        T\frac{\rho_2^2\varphi_1^{-1}(\tau)}{t^{*}\rho_2^2+(T-t^{*})\rho_1^2}\, , & \tau\in [0,t^{*}]  \\
        T\frac{\rho_2^2t^{*}+\rho_1^2(\varphi_2^{-1}(\tau)-t^{*})}{t^{*}\rho_2^2+(T-t^{*})\rho_1^2}\, , & \tau\in [t^{*},T]\, .
    \end{array}\right.
\end{equation}

For a process on the two-dimensional plane, the mapping $\Omega^{\mathbb{R}^2}$ and its inverse have the following form. The coordinates $\rho_1,\varphi_1,\rho_2,\varphi_2$ are related to $\rho,\varphi$ by eqs. (\ref{rho12})-(\ref{rhophi}). The angular coordinates $\psi_1$, $\psi_2$, $\alpha_1$, $\alpha_2$ are expressed in terms of $\psi$, $\alpha$ as
\begin{equation}
    \alpha_1=\alpha\, ,\qquad \alpha_2=\alpha_1+\psi(t^{*})\, ,
\end{equation}
\begin{equation}
    \psi_1(\tau)=\psi(\tau) \, ,\quad \tau\in[0,t^{*}]\, ,
\end{equation}
\begin{equation*}
    \psi_2(\tau)=\psi(\tau)-\psi(t^{*})\, ,\qquad \tau\in [t^{*},T]\, .
\end{equation*}
Conversely, given $\psi_1$, $\psi_2$, $\alpha_1$, $\alpha_2$, one can uniquely reconstruct $\psi,\alpha$
\begin{equation}
    \alpha=\alpha_1\, ,\qquad \psi(\tau)=\left\{\begin{array}{cc}
        \psi_1(\tau) & \tau\in[0,t^{*}] \\
        \psi_2(\tau)+\alpha_2-\alpha_1 & \tau\in[t^{*},T]\, .
    \end{array}\right.
\end{equation}

In the case of trajectories on the cone, the coordinates $\rho_1,\varphi_1,\rho_2,\varphi_2$ are related to $\rho,\varphi$ using formulas (\ref{rho12})-(\ref{rhophi}). The angular coordinates $\psi,\psi_1,\psi_2$ are related by
\begin{equation}
    \psi(\tau)=\left\{\begin{array}{cc}
        \psi_1(\tau), & \tau\in[0,t^{*}] \\
        \psi_2(\tau), & \tau\in[t^{*},T]\, ,
    \end{array}\right.
\end{equation}
which defines $\Omega^{\mathrm{Cone}}$.

\section{Appendix B}
We show the equality of eqs. (\ref{wtilde}) and (\ref{tildewxi}). Using the expression (\ref{mu}) for the quasi-invariant measure on the group of diffeomorphisms and the expression (\ref{win1d}) for the Wiener measure, we represent $\tilde w_{\sigma}$ in the form
\begin{equation}
     \tilde w_{\sigma}(d\xi)= \rho\exp\left(-\frac{\sigma^2}{4\rho^2}\right)\sqrt{\dot \varphi(0)\dot \varphi(T)} \times
\end{equation}
\begin{equation*}
   \exp\left(\frac{\rho^2}{4\sigma^2}\left[\frac{\ddot \varphi(0)}{\dot\varphi(0)}-\frac{\ddot \varphi (T)}{\dot \varphi(T)}\right]+\frac{\rho^2}{4\sigma^2}\int_0^Tds\left\{\frac{d}{ds}\left(\frac{\ddot \varphi(s)}{\dot \varphi(s)}\right)-\frac12\left(\frac{\ddot \varphi(s)}{\dot \varphi(s)}\right)^2\right\}\right)\times
\end{equation*}
\begin{equation*}
\times\exp\left(-\frac{\rho^2}{2\sigma^2}\int_0^Tds\, (\dot \psi(s))^2\right)d\varphi\, d\psi\, d\rho\, .
\end{equation*}
After simplifying the expression in the exponent, we obtain
\begin{equation}\label{tildesmp}
    =\rho\sqrt{\varphi'(0)\varphi'(T)}\exp\left(-\frac{\sigma^2}{4\rho^2}\right)\exp\left(-\frac{1}{2\sigma^2}\int_0^Tds\left\{\frac{\rho^2}{4}\left(\frac{\ddot \varphi(s)}{\dot \varphi(s)}\right)^2+\rho^2(\dot \psi(s))^2\right\}\right)\, .
\end{equation}
We change the integration variable $s$ to the variable $\tau$
\begin{equation}
s=\varphi^{-1}(\tau)\, .
\end{equation}
For the derivatives we have
\begin{equation}
    \dot \varphi(s)=\frac{1}{\dot \varphi^{-1}(\tau)}\, ,\qquad \ddot \varphi(s)=\frac{1}{\dot \varphi^{-1}(\tau)}\frac{d}{d\tau}\left(\frac{1}{\dot \varphi^{-1}(\tau)}\right)=-\frac{\ddot \varphi^{-1}(\tau)}{(\dot \varphi^{-1}(\tau))^3}\, .
\end{equation}
For the integral in (\ref{tildesmp}) we obtain the following equality
\begin{equation}\label{intint}
    \int_0^Tds\left\{\frac{\rho^2}{4}\left(\frac{\ddot \varphi(s)}{\dot \varphi(s)}\right)^2+\rho^2(\dot \psi(s))^2\right\}=\int_0^T d\tau \left\{\frac{\rho^2}{4}\frac{(\ddot \varphi^{-1}(\tau))^2}{(\dot \varphi^{-1}(\tau))^3}+\rho^2\dot \varphi^{-1}(\tau)(\dot \psi(\varphi^{-1}(\tau)))^2\right\}
\end{equation}
We use (\ref{decompcone}) and express $\rho,\varphi$ in terms of $\xi^0,\xi^1$
\begin{equation}
    \dot \varphi^{-1}(\tau)=\rho^2\frac{1}{(\xi^0(\tau))^2-(\xi^1(\tau))^2}\, ,\qquad \ddot \varphi^{-1}(\tau)=-2\rho^2\frac{\xi^0(\tau)\dot \xi^0(\tau)-\xi^1(\tau)\dot \xi^1(\tau)}{((\xi^0(\tau))^2-(\xi^1(\tau))^2)^2}\, .
\end{equation}
For the first term on the right-hand side of (\ref{intint}) we will have
\begin{equation}\label{dr2}
    \frac{\rho^2}{4}\frac{(\ddot \varphi^{-1}(\tau))^2}{(\dot \varphi^{-1}(\tau))^3}=\frac{(\xi^0(\tau)\dot \xi^0(\tau)-\xi^1(\tau)\dot \xi^1(\tau))^2}{(\xi^0(\tau))^2-(\xi^1(\tau))^2}\, .
\end{equation}
From (\ref{decompcone}) for $\psi(s)$ we will have
\begin{equation}
    \dot \psi(s)=\frac{d}{ds}\left(\mathrm{arctanh}\left(\frac{\xi^1(\varphi(s))}{\xi^0(\varphi(s))}\right)\right)=\frac{\dot \xi^1(\varphi(s))\xi^0(\varphi(s))-\dot \xi^0(\varphi(s))\xi^1(\varphi(s))}{(\xi^0(\varphi(s)))^2-(\xi^1(\varphi(s)))^2}\dot \varphi(s)\, ,
\end{equation}
\begin{equation}
    \dot \psi(\varphi^{-1}(\tau))=\frac{\dot \xi^1(\tau)\xi^0(\tau)-\dot \xi^0(\tau) \xi^1(\tau)}{(\xi^0(\tau))^2-(\xi^1(\tau))^2}\frac{1}{\dot \varphi^{-1}(\tau)}\, .
\end{equation}
The second term in (\ref{intint}) takes the form
\begin{equation}\label{r2dpsi2}
    \rho^2\dot \varphi^{-1}(\tau)(\dot \psi(\varphi^{-1}(\tau)))^2=\frac{(\dot \xi^1(\tau)\xi^0(\tau)-\dot \xi^0(\tau) \xi^1(\tau))^2}{(\xi^0(\tau))^2-(\xi^1(\tau))^2}\, .
\end{equation}
Combining (\ref{dr2}) and (\ref{r2dpsi2}), we obtain
\begin{equation}
    \int_0^T d\tau \left\{\frac{\rho^2}{4}\frac{(\ddot \varphi^{-1}(\tau))^2}{(\dot \varphi^{-1}(\tau))^3}+\rho^2\dot \varphi^{-1}(\tau)(\dot \psi(\varphi^{-1}(\tau)))^2\right\}=
\end{equation}
\begin{equation*}
    =\int_0^Td\tau\left\{\frac{(\xi^0(\tau)\dot \xi^0(\tau)-\xi^1(\tau)\dot \xi^1(\tau))^2}{(\xi^0(\tau))^2-(\xi^1(\tau))^2}+\frac{(\dot \xi^1(\tau)\xi^0(\tau)-\dot \xi^0(\tau) \xi^1(\tau))^2}{(\xi^0(\tau))^2-(\xi^1(\tau))^2}\right\}=
\end{equation*}
\begin{equation*}
=\int_0^T d\tau \left\{(\dot \xi^0)^2-(\dot \xi^1)^2+2\frac{(\xi^0\dot \xi^1-\xi^1\dot \xi^0)^2}{(\xi^0)^2-(\xi^1)^2}\right\}\, .
\end{equation*}

\end{document}